\documentclass[fp,twocolumn]{jpsj3}
\usepackage{txfonts}
\usepackage{algorithm}
\usepackage{algorithmic}
\usepackage{amsmath}
\usepackage{multirow}

\usepackage{bm}
\usepackage{color}


\title{Derivation of QUBO formulations for sparse estimation}

\author{Tomohiro Yokota$^1$
  , Makiko Konoshima$^2$, Hirotaka Tamura$^2$, Jun Ohkubo$^{1,3}$}
\inst{$^1$Graduate School of Science and Engineering, Saitama University,
  255 Shimo-Okubo, Sakura-ku, Saitama-shi, 338-8570, Japan\\
  $^2$Fujitsu Laboratories Ltd.,
  4-1-1 Kawasaki, Kanagawa 211-8558, Japan\\
  $^3$JST, PREST, 4-1-8 Honcho, Kawaguchi, Saitama 332-0012, Japan} 

\abst{
We propose a quadratic unconstrained binary optimization (QUBO) formulation of the $\ell_{1}$-norm, which enables us to perform sparse estimation of Ising-type annealing methods such as quantum annealing. The QUBO formulation is derived using the Legendre transformation and the Wolfe theorem, which have recently been employed to derive the QUBO formulations of ReLU-type functions. It is shown that a simple application of the derivation method to the $\ell_{l}$-norm case results in a redundant variable. Finally a simplified QUBO formulation is obtained by removing the redundant variable.
}

\begin{document}
\maketitle

\section{Introduction}

In recent years, some novel computing hardwares have been developed and implemented; 
for example, there are some Ising-type annealing machines such as ``D-Wave 2000'' by the Canadian company D-Wave \cite{d-wave01,d-wave02} and ``Fujitsu's Digital Annealer'' by the Japanese company Fujitsu, which is a physically inspired annealing accelerator based on classical (i.e., non-quantum) logic circuits \cite{DA}.
Annealing machines are used to obtain approximate solutions for optimization problems.
Optimization problems play important roles in various research areas such as data mining and machine learning.
In particular, the quantum annealing method, which was originally proposed by Kadowaki and Nishimori\cite{Kadowaki1998}, and a similar idea called adiabatic quantum computing \cite{Farhi2001} have attracted considerable attention, and recently, 
several researches were conducted on its practical applications (for example, see the paper by Tanahashi et al.\cite{Tanahashi2019})
Biamonte et al. discussed the use of quantum Boltzmann machines for machine learning\cite{Biamonte}, and there are many challenging tasks from the viewpoints of hardware and software. 
Although the small size of the system poses a restriction, the number of available qubits (or classical bits) has been increasing each year, which enables us to tackle practical and large optimization problems.

The above annealing hardwares require quadratic unconstrained binary optimization (QUBO) formulations; these hardwares are based on the Ising-type Hamiltonian, and hence it is necessary to convert the original cost functions in optimization problems into QUBO formulations
(the QUBO formulation is equivalent to the Ising model).
Although continuous variables can be expressed as Ising-type variables via adequate binary expansions, 
it is not straightforward in general to reformulate the original cost functions as QUBO formulations.
Some reformulations were given in the review paper by Lucas \cite{Lucas2014}, 
and it was shown that logic gates were expressed in the form of QUBO formulations\cite{Whitfield}.
However, a systematic method to derive the QUBO formulations has not been found yet.
Recently, the Legendre transformation was employed to derive the QUBO formulation of the $q$-loss function \cite{q-loss}; 
the $q$-loss function was proposed as a cost function with robust characteristics against the label noise in machine learning. 
The derivation technique based on the Legendre transformation revealed that some mathematical transformation would be needed to convert some types of cost functions into QUBO formulations.
It has been clarified that the Legendre transformation is not enough to deal with Rectified Linear-Unit (ReLU) type functions \cite{relu};
the Wolfe duality theorem \cite{wolfe} was employed to derive the QUBO formulations for ReLU-type functions.
These works also showed that the derivation of QUBO formulations was not straightforward and that we need further and careful considerations, which depend on the original cost functions.

As shown above, there have been some works on deriving the QUBO formulations for machine learning problems.
Clearly, there are many other research fields involving optimization problems, and one of them is data analysis (data mining).
It is known that the concept of ``regularization'' plays an important role in data analysis and machine learning.
For example, the $\ell_{2}$ norm is widely used; a linear regression with $\ell_{2}$ regularization is called ridge regression, and it is widely used in various practical applications.
In addition, the $\ell_{1}$ norm is another useful example of regularization, which is used to introduce a kind of ``sparseness'' for the solution; sparse estimation is one of the hot topics in the field of data analysis.
The least absolute shrinkage and selection operator (LASSO) \cite{lasso} is a well-known practical method to achieve sparse estimations, in which the $\ell_{1}$ norm is added to a least-squares cost function.
Recently, the idea of sparse estimation was applied to black hole analysis \cite{black-hole}.
The data size for a black hole is small, and hence it is difficult to observe the image of a black hole directly because of the low resolution of images.
Therefore, simultaneous measurements were taken from radio telescopes all over the world,
and the method based on sparse estimation was applied to the observed big data in order to extract only the essential information; finally, the imaging of the black holes was achieved.
Note that the $\ell_{2}$ norm is simply connected to the QUBO formulation because of the quadratic form; 
in contrast, the $\ell_{1}$ norm has a non-differentiable point, and the QUBO formulation has not been derived yet.

In order to include ``sparseness'' into the formulation, $\ell_{0}$ norms are more preferable.
Although it is difficult to treat $\ell_{0}$ norms in general because of the non-convex characteristics, $\ell_{0}$ norms have been used for the QBoost algorithm\cite{Neven2012}, which is a kind of boosting algorithm suitable for Ising-type hardware. As discussed in Ref.~\citen{Neven2012}, when we discretize and approximate the components in a weight vector as binary ($0$ or $1$) variables, the $\ell_{0}$ norm of the weight vector is given simply as a summation of the binary variables; this discretization is preferable for Ising-type hardware. However, since it is still difficult to deal with the $\ell_{0}$ norm and the $\ell_{1}$ norm has been widely used in the regularization for the sparseness characteristics, here we focus on the $\ell_{1}$ norm and its QUBO formulation.

In this work, we derive the QUBO formulation of the $\ell_{1}$ norm.
In order to obtain the QUBO formulation, both the Legendre transformation and the Wolfe duality theorem are employed.
Furthermore, it is clarified that only simple applications of the previous derivation techniques are not enough;
through numerical checks and reconsideration for the derived formulation, a simplified QUBO formulation is finally derived.
The simplified QUBO formulation has a smaller number of variables than the naive derived formulation;
such reduction in the number of variables is important for hardware implementation 
because the number of qubits (or classical bits) in the current Ising-type hardwares are restricted.

This paper is organized as follows.
Section~2 explains the QUBO formulation and discusses the related previous works.
The important techniques used for the derivations are also given for later use.
In Sect.~3, the QUBO formulation of the $\ell_{1}$ norm is derived,
and the numerical checks are described.
Section~4 gives the main result of this paper;
a simplified version of the QUBO form is given,
in which a variable is removed from the naive QUBO formulation derived in Sect.~3.
Section 5 provides the concluding remarks and mentions the future works.

\section{Backgrounds and preliminaries}

The aim of this work is to derive the QUBO formulation of the $\ell_1$ norm-type function.
As mentioned in the Introduction, the QUBO formulation of a slightly complicated function, namely the $q$-loss function,
has already been derived\cite{q-loss}.
One may think that a combination of $q$-loss functions could be used as regularization functions;
however the $q$-loss function is not enough to obtain the $\ell_1$ norm.
Based on the derivation techniques used in the $q$-loss case,
the QUBO formulation of the ReLU-type function was discussed in Ref.~\citen{relu}.
In Ref.~\citen{relu}, two techniques, namely the Legendre transformation and the Wolfe duality theorem, were employed,
which also play important roles in our discussions.
In this section, we briefly provide some background knowledge and discuss the previous works,
starting from a brief explanation of the QUBO formulations and Ising model.
In particular, the Legendre transformation and the Wolfe duality theorem are briefly described.

\subsection{QUBO and Ising model}

As mentioned in the Introduction, Ising-type annealing machines require the Ising Hamiltonian or the QUBO formulation in order to solve combinatorial optimization problems. 
The QUBO form involves binary variables, which take value of only $1$ or $0$.
The $0$-$1$ binary variables are sometimes suitable for considering combinatorial optimizations; 
the binary expansions of continuous variables naturally introduce binary expressions.
Since the QUBO formulation and Ising model are equivalent, it is possible to convert the QUBO form into the Ising model, and vice versa.
The Ising model is represented as follows:
\begin{align}
  H=-\sum_{i,j}{J_{ij}\sigma_{i}\sigma_{j}}-\sum_{i}{h_{i}\sigma_{i}},
\end{align}
where $\sigma_{i} \in \{-1, +1\}$ is a spin variable for spin $i$, $J_{ij} \in \mathbb{R}$ is a coefficient related to the quadratic term between spins $i$ and $j$, and $h_{i} \in \mathbb{R}$ is a coefficient for the linear term with spin $i$. 
Let $q_{i} \in \{0,1\}$ be a binary variable corresponding to the $i$-th spin.
Then by applying the variable transformation $q_{i} = (\sigma_{i}+1)/2$,
we have
\begin{align}
  H=-\sum_{i,j}{\widetilde{J}_{i,j}q_{i}q_{j}}-\sum_{i}{\widetilde{h}_{i}q_{i}},
\end{align}
where $\widetilde{J}_{i,j}$ and $\widetilde{h}_{i}$ are transformed from $\{J_{ij}\}$ and $\{h_{i}\}$ adequately.
As for the relations between the QUBO formulation and the Ising Hamiltonian, please see, for example, Ref.~\citen{Tanahashi2019};
in Ref.~\citen{Tanahashi2019}, some examples of the QUBO formulations for typical optimization problems are also given.

\subsection{Legendre transformation}
\label{sec:Legendre}

For the convenience of the readers, here we give a brief notation of the Legendre transformation.

If a function $f_{L}$ is convex, the Legendre transformation of $f_{L}$, the so-called conjugate function of $f_{L}$, is given as follows:
\begin{align}
\label{eq:Legmax}
f_{L}^{*}(t)=\sup_{x}\{t x - f_{L}(x)\}.
\end{align}
That is, the variable $t$ is introduced, and the function of $x$ is transformed into a function of $t$.
In addition, \eqref{eq:Legmax} is equivalent to following equation:
\begin{align}
\label{eq:Legmin}
f_{L}^{*}(t)=-\inf_{x}\{f_{L}(x) - t x\}.
\end{align}

\subsection{Previous work 1: $q$-loss function}

Here, we briefly review the previous work by Denchev et al.\cite{q-loss}
The following $q$-loss function was proposed in Ref.~\citen{q-loss}:
\begin{align}
  L_{q}(m)=\min{[(1-q)^{2}, (\max{[0,1-m]})^{2}]},
\label{q-loss_function}
\end{align}
where $q \in (-\infty,0]$ is a parameter and $m$ is a continuous variable. 
It also includes a discussion on the application of the $q$-loss function in machine learning problems, and it was clarified that the $q$-loss function has a robust feature against label noise.
Since \eqref{q-loss_function} has a $\max$ function, 
it would not be easy to see the QUBO formulation of the $q$-loss function.
Denchev et al. employed the Legendre transformation and derived the following function\cite{q-loss}:
\begin{align}
  L_{q}(m)=\min_{t}{\left\{(m-t)^{2}+(1-q)^{2}\frac{(1-\text{sign}(t-1))}{2}\right\}}, 
\label{q-loss_function_legendre}
\end{align}
where $t$ is an additional variable introduced via the Legendre transformation.
Although the variables $m$ and $t$ in \eqref{q-loss_function_legendre} are continuous, 
the use of binary expansions gives the QUBO formulation for the $q$-loss function.
As for details of the binary expansions, please see Ref.~\citen{q-loss}.
Note that the sign function in \eqref{q-loss_function_legendre} is also expressed as a one-body term when we employ the binary expansion.

\subsection{Wolfe-duality}
\label{sec:wolfe}

In nonlinear programming and mathematical optimization, the Wolfe duality theorem \cite{wolfe} is used to convert the main problem with inequality constraints into a dual problem.
For a differentiable objective function and differentiable constraints, the main problem is written as follows:
\begin{align}
\begin{cases}
    \, \text{minimize}_{\bm{x}}  \quad  f_{\mathrm{W}}(\bm{x}) \quad \quad \ (\bm{x} \in \mathbb{R}^{n}),\\
    \, \text{subject to}  \ \quad h_{i}(\bm{x})\leq 0 \quad (i=1,2,\dots,l),
  \label{object_function}
\end{cases}
\end{align}
where $f_{\mathrm{W}}(\bm{x})$ is the convex function to be optimized and $\{h_{i}(\bm{x})\}$ are convex inequality constraints. The Lagrangian function for this optimization problem is
\begin{align}
  L(\bm{x},\bm{z})=f_{\mathrm{W}}(\bm{x})+\bm{z}^{T}h(\bm{x}),
\end{align}
where $\bm{z}$ is a vector of the Legendre coefficients. 
Then, according to the Wolfe dual theorem, the minimization problem in Eq.~\eqref{object_function} is equivalent to the following maximization problem:
\begin{align}
\begin{cases}
    \, \text{maximize}_{\bm{x},\bm{z}}  \quad L(\bm{x},\bm{z}) \quad \quad \quad \ ((\bm{x},\bm{z})\in \mathbb{R}^{n}\times\mathbb{R}^{l}),\\
    \, \text{subject to}  \qquad \nabla L(\bm{x},\bm{z})=0 \quad (\bm{z} \geq 0).
\end{cases}
\end{align}
As shown above, the Wolfe dual theorem transforms the minimization problem into a maximization problem.

\subsection{Previous work 2: ReLU-type function}
\label{sec:ReLU}

In Ref.~\citen{relu}, the QUBO form of the following ReLU-type function was discussed:
\begin{align}
  f_{\mathrm{ReLU}}(m)=-\min{(0,m)}. 
\label{ReLU_function}
\end{align}
Note that the function $f_{\mathrm{ReLU}}(m)$ becomes a conventional ReLU function when the variable transformation $m \to -m$ is employed.
As shown in Ref.~\citen{relu}, a naive application of the Legendre transformation to the function $f_{\mathrm{ReLU}}(m)$ in \eqref{ReLU_function} gives the following expression:
\begin{align}
  f_{\mathrm{ReLU}}(m)=-\min_{t}{\{-mt\}} \quad \text{subject to} \quad -1\leq t\leq 0, 
\label{ReLU_function_legendre}
\end{align}
where $t$ is a new variable stemming from the Legendre transformation.

Although \eqref{ReLU_function_legendre} includes the QUBO formulation after adequate binary expansions, it is not suitable for optimization problems.
This is because of the minus sign before the $\min$ function;
when the ReLU-type function is used as a constraint or penalty term for an optimization problem with a cost function $C(m)$, the overall minimization problem is, for example, given as follows:
\begin{align}
  \min_{m} \left\{ C(m)+f_{\mathrm{ReLU}}(m) \right\} &= \min_{m}\left\{ C(m)-\min_{t} \{-mt\} \right\} \nonumber \\
  &\neq \min_{m, t} \left\{C(m)-(-mt)\right\} ,
\end{align}
and hence the cost function $C(m)$ and the ReLU-type function $f_{\mathrm{ReLU}}(m)$ cannot be minimized simultaneously.
Therefore, the Wolfe duality theorem was employed in Ref.~\citen{relu}, and finally the following formulation was derived:
\begin{align}
  f_{\mathrm{ReLU}}(m) = \min_{t, z_{1}, z_{2}} \left\{mt+z_{1}(t+1)-z_{2}t-M(-m-z_{1}+z_{2})^{2}\right\},
\label{ReLU_function_wolfe}
\end{align}
where $M$ is a large positive constant. 
It is easy to see that \eqref{ReLU_function_wolfe} can be used in combination with the cost function $C(m)$.

Note that it is possible to obtain the $\ell_1$ norm from a combination of two ReLU-type functions.
However, this construction results in redundant variables.
In the next section, we will directly derive the QUBO formulation starting from the $\ell_1$ norm.
In the derivation, the two techniques explained in Sect.~\ref{sec:Legendre} and Sect.~\ref{sec:wolfe} play important roles.

\section{Naive derivation of QUBO formulation for $\ell_{1}$-norm} \label{Native_derivation}

In this section, the Legendre transformation and the Wolfe dual theorem are applied to the $\ell_{1}$ norm-type function naively. 

\subsection{QUBO formulation}

\begin{figure}[tb]
  \begin{center}
    \includegraphics[keepaspectratio,width=60mm]{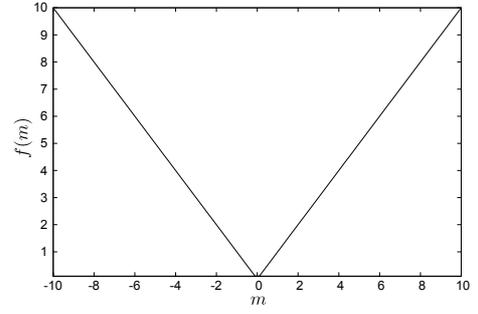}
    \caption{The function form of $\ell_{1}$-norm in \eqref{l1-norm}.}
    \label{fig:absolute}
  \end{center}
\end{figure}

Although the $\ell_{1}$-norm is usually denoted as the absolute value of a variable, i.e., $\lvert m \rvert$, here we employ the following function $f(m)$:
\begin{align}
  f(m)=-\min{\{-m,m\}}.
 \label{l1-norm}
\end{align}
Note that $f(m)$ can be expressed as follows:
\begin{align}
f(m) &= -\min{\{0,m\}} - \min\{-m,0 \} \nonumber \\
&= f_{\mathrm{ReLU}}(m) + f_{\mathrm{ReLU}}(-m),
\end{align}
where $f_{\mathrm{ReLU}}(m)$ is the ReLU-type function in \eqref{ReLU_function}.
Hence, it is easy to obtain the QUBO formulation for $f(m)$ based on the discussion in Sect.~\ref{sec:ReLU}.
However, the QUBO formulation needs six additional variables.
The derivation below enables us to obtain the QUBO formulation for $f(m)$ with only three additional variables.

Here, we try a naive application of the Legendre transformation to the function $f(m)$ in \eqref{l1-norm}. 
In order to employ the Legendre transformation, 
we use the following form of \eqref{l1-norm}:
\begin{align}
  f(m)=
\begin{cases}
   \, -m\quad & (m <0), \nonumber \\
   \,  m\quad & (m\geq 0). 
\end{cases}
\end{align}
Then, the Legendre transformation in \eqref{eq:Legmin} is performed for each domain as follows:
\begin{itemize}
\item[(a)] $m<0$: \\
  The gradient in the domain is always $-1$. Hence, the conjugate function is
  \begin{align}
    f^{*}(t)=-\inf_{m} \{-m - mt\}=-\inf_{m}\{-m(1+t)\} = 0. \nonumber
  \end{align}
  In addition, the only possible value of $t$ is $t = -1$.
\item[(b)] $m=0$: \\
  Since the left derivative at this point is $f_{-}'(m) = -1$ and the right derivative is $f_{+}'(m) = 1$, 
the gradient value takes an arbitrary value within $-1$ to $1$. 
Hence, the conjugate function is $f^{*}(t) = 0$ with the domain $t \in [-1,1]$.
\item[(c)] $m>0$: \\
  The gradient in the domain is always $1$. Hence, the conjugate function is
  \begin{align}
    f^{*}(t)=-\inf_{m} \{m - mt\}=-\inf_{m}\{-m(-1+t)\} = 0. \nonumber
  \end{align}
  In addition, the only possible value of $t$ is $t = 1$.
\end{itemize}

From the above discussion, the conjugate function of $f(m)$ is $f^{*}(t) = 0\ (-1 \leq t \leq 1)$. 
When we apply the Legendre transformation to $f^{*}(m)$ again, 
$f(m)$ is adequately recovered since the function $f(m)$ is convex. 
Therefore, we find the quadratic form of $f(m)$ as follows:
\begin{align}
  F(m) = -\min_{t}{\{-mt\}} \quad \text{subject to} \quad \-1 \leq t \leq 1.
\label{legendre}
\end{align}
In order to emphasize the fact that it is the quadratic form of $f(m)$, we introduce $F(m)$ instead of $f(m)$.

As shown in Sect.~\ref{sec:ReLU}, although the expression obtained via the Legendre transformation has a quadratic form, it cannot be combined with another cost function.
Hence, the Wolfe dual theorem is employed;
the following expression is immediately obtained by applying the Wolfe dual theorem to $F(m)$:
\begin{align}
  &\widetilde{F}(m)=\max_{t,z_{1},z_{2}} \{-mt-z_{1}(t+1)+z_{2}(t-1)\} \label{wolf} \\
  & \text{subject to} \quad 
 \begin{cases}
   \, -m-z_{1}+z_{2}=0, \nonumber \\
   \ -1\leq t\leq 1, \, 0 \leq z_{1}, \, 0 \leq z_{2}. \\
 \end{cases}
\end{align}
This reformulation has the equality constraint, $-m-z_{1}+z_{2}=0$; 
in order to embed this constraint into the QUBO formulation, it is enough to add the squared term as a penalty. 
Therefore, the optimization problem \eqref{wolf} can be represented as follows:
\begin{align}
    &\widetilde{F}(m) = \min_{t,z_{1},z_{2}}{\{mt+z_{1}(t+1)-z_{2}(t-1)} \nonumber \\
    &\qquad\qquad\qquad +M(-m-z_{1}+z_{2})^{2}\} \nonumber  \\
  & \text{subject to} \quad -1\leq t\leq 1, \, 0 \leq z_{1}, 0 \leq z_{2}. \label{after_wolf}
\end{align}
where $M$ is a constant that takes a large value to ensure that the equality constraint, $-m-z_{1} + z_{2} = 0$, is satisfied.
Note that there are some remaining inequality constraints, namely $-1\leq t\leq 1, 0 \leq z_{1}$, and $0 \leq z_{2}$;
these inequality constraints can be easily realized by expanding the variables $t, z_{1}$, and $z_{2}$ in the binary expressions that satisfy the corresponding domain constraints.

As a result, the QUBO formulation for the $\ell_{1}$-norm is expressed using three additional variables.

\subsection{Numerical validation}

Here, we discuss the numerical checks that are performed to confirm the validity of the obtained QUBO formulation in \eqref{after_wolf}. 
We expect that the QUBO formulation will be used with quantum annealing methods or simulated annealing methods; 
here, simulated annealing algorithms are employed.
Since the purpose here is to verify the obtained QUBO formulation, 
the function in \eqref{after_wolf} is experimented with continuous variables without binary expansions. 

The aim here is to check whether $\widetilde{F}(m)$ in \eqref{after_wolf} gives the $\ell_{1}$-norm in \eqref{l1-norm} or not. 
We randomly generate $m$ and check the value of $\widetilde{F}(m)$ by using a simulated annealing method for continuous variables. 
Here, $m$ is chosen from a uniform distribution with a range of $[-10,10]$.
A simulated annealing is performed for each chosen $m$.
In each numerical experiment, the initial conditions for the additional variables, 
$t, z_{1}$, and $z_{2}$ are chosen as follows:
\begin{itemize}
\item $t$ is generated from a uniform distribution with a range of $[-1,1]$.
\item $z_{1}$ and $z_{2}$ are generated from a uniform distribution with a range of $[0,10]$.
\end{itemize}

As the simulated annealing method for continuous variables, 
we employ a conventional Metropolis-Hastings-type method.
In order to generate the subsequent candidates of a state, 
each variable moves by an amount of $+0.001$ or $-0.001$ with the same probability for each iteration.
At each iteration, the temperature is changed with the annealing schedule: $T_{n+1}=0.9999T_{n}$, 
where $n$ is the iteration step.
The initial temperature is set as $T_{1}=1000$, 
and the annealing is stopped when the temperature becomes lower than $10^{-3}$.

\begin{figure}[tb]
  \begin{center}
    \includegraphics[keepaspectratio,width=80mm]{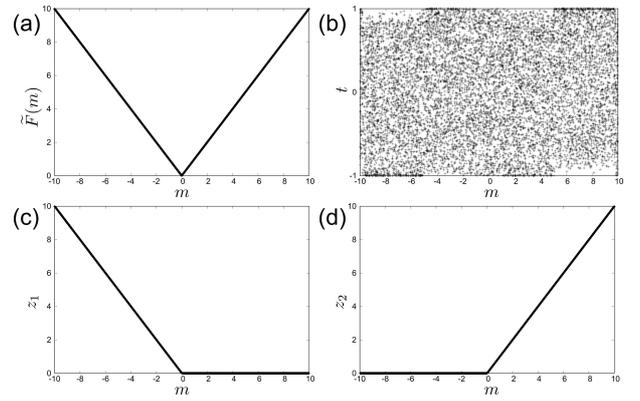}
    \caption{Numerical results obtained from the annealing method. Here, $m$ is chosen randomly, and the annealing is performed. (a) Numerical results for the derived formulation in \eqref{after_wolf}; we confirm that the annealing on \eqref{after_wolf} gives the original shape of the $\ell_1$ function. (b), (c), and (d) correspond to $t$, $z_{1}$, and $z_{2}$ at the optimum states in the annealing, respectively.}
    \label{fig:minimum1}
  \end{center}
\end{figure}

Figure~\ref{fig:minimum1} shows the results of the annealing;
Fig.~\ref{fig:minimum1}(a) shows that the $\ell_{1}$-norm is adequately recovered by the optimization of \eqref{after_wolf}. 
Figs.~\ref{fig:minimum1}(b), (c), and (d) show the values of the three additional variables. It is clear that $z_{1}$ and $z_{2}$ converge to specific values, but $t$ takes various values randomly. This means that $t$ could not be necessary for the optimization, which may give us a further simplified QUBO formulation for the $\ell_{1}$-norm.

\section{Reduced QUBO formulation}
\subsection{Reduction of the variable in the Legendre transformation}

As discussed above, the naive application of the Legendre transformation and the Wolfe dual theorem gives a QUBO formulation with three additional variables.
However, from the numerical experiments in the previous section, it is revealed that the variable $t$, which stems from the Legendre transformation, may not be necessary for the optimization problem.
Because of the restriction with regard to the number of spin variables, it is preferable to have smaller number of variables in general.
Hence, here we try to achieve further reduction in the number of variables from that in the QUBO formulation in \eqref{after_wolf}. 

In order to eliminate $t$ from \eqref{after_wolf}, 
we focus on the equality constraint $-m-z_{1} + z_{2} = 0$.
By employing the equality $z_{2} = m+z_{1}$, we have 
\begin{align}
  \widetilde{F}(m)&=\min_{t,z_{1},z_{2}}{\{mt+z_{1}(t+1)-z_{2}(t-1)} \nonumber \\
  &\quad\quad\quad\,+M(-m-z_{1}+z_{2})^{2}\} \nonumber \\
  &=\min_{t,z_{1},z_{2}}{\{mt+z_{1}(t+1)-(m+z_{1})(t-1)} \nonumber \\
  &\quad\quad\quad\,+M(-m-z_{1}+z_{2})^{2}\} \nonumber \\
  &=\min_{z_{1},z_{2}}{\{z_{1}+(m+z_{1})+M(-m-z_{1}+z_{2})^{2}\}} \nonumber \\
  &=\min_{z_{1},z_{2}}{\{z_{1}+z_{2}+M(-m-z_{1}+z_{2})^{2}\}}. 
\end{align}
Finally we obtain the following simplified QUBO formulation for the $\ell_{1}$-norm:
\begin{align}
&\widehat{F}(m) = \min_{z_{1},z_{2}}{\{z_{1}+z_{2}+M(-m-z_{1}+z_{2})^{2}\}} \nonumber \\
  & \text{subject to} \quad  0 \leq z_{1}, 0 \leq z_{2}. \label{review_formulation}
\end{align}
where the new expression $\widehat{F}(m)$ is introduced to clarify the difference from ~\eqref{after_wolf}.
This conversion from \eqref{after_wolf} to \eqref{review_formulation} is possible because the penalty term, $M(-m-z_{1}+z_{2})^{2}$, forces the equality constraint to be satisfied.

\subsection{Numerical validation}

In order to check the validity of the simplified QUBO formulation in \eqref{review_formulation}, we again perform numerical experiments. 
The same settings and procedures as those in the previous section are employed for the annealing, except for the exclusion of the variable $t$.
Consequently, we obtained the same figures as those in Fig.~\ref{fig:minimum1}, except for the figure with respect to $t$
(the top-right figure in Fig.~\ref{fig:minimum1}).
Without using the redundant variable $t$, we completely recovered the $\ell_1$ function.
This shows that the simplified QUBO formulation works well.

\section{Concluding remarks}

In this work, the QUBO formulation for the $\ell_{1}$-norm was derived using the Legendre transformation and the Wolfe theorem.
It was shown that one of the variables introduced by the derivation is redundant,
and it was numerically confirmed that the final simplified QUBO formulation obtained by excluding this variable works well.
At first glance, it would be difficult to see the connection between the $\ell_{1}$-norm and the final expression in \eqref{review_formulation};
this nontrivial result is the main contribution of the present work.

There are some points that need to be noted regarding the derivation of the QUBO formulations.
As discussed in Sect.~1, there is no systematic way to derive the QUBO formulation.
While the Legendre transformation and the Wolfe theorem can be used in many cases,
the derivation generally results in additional variables;
the reduction of additional variables is very important for Ising-type hardware because of the current limitation with regard to spin variables.
As shown in the derivation, the variable $t$, which was introduced by the Legendre transformation, was finally excluded. It is still not clear whether the use of the Legendre transformation is necessary or not; at this stage, the procedure (the Legendre transformation $\to$ the Wolfe dual theorem $\to$ reduction of variables) is straightforward and understandable.
Clearly, there could be more suitable derivation methods.
In addition, when the $\ell_{1}$-norm is combined with another cost function to perform sparse estimation with Ising-type hardware, it is necessary to add two variables for each estimated value,
which means that the number of additional variables is still large. 
In future works, we need to find methods that can achieve further reduction in practical problems with large size.







\end{document}